# Demonstration of a strain-mediated magnetoelectric write and read unit in a $Co_{60}Fe_{20}B_{20}$/ $Pb(Mg_{1/3}Nb_{2/3})_{0.7}Ti_{0.3}O_3$ heterostructure

Tingting Shen [‡ 1,3], Vaibhav Ostwal[‡2,3], Kerem Y. Camsari[2], Joerg Appenzeller[2,3]

1. Department of Physics and Astronomy, Purdue University, West Lafayette, Indiana 47907, United States

2. Department of Electrical and Computer Engineering, Purdue University, West Lafayette, Indiana 47907, United States

3. Birck Nanotechnology Center, Purdue University, West Lafayette, Indiana 47907, United States

[‡]. Authors contributed equally to the article

## Abstract

Taking advantage of the Magnetoelectric (ME) and its inverse effect, this article demonstrates strain-mediated magnetoelectric write and read operations simultaneously in $Co_{60}Fe_{20}B_{20}$/ $Pb(Mg_{1/3}Nb_{2/3})_{0.7}Ti_{0.3}O_3$ heterostructures without using any symmetry breaking magnetic field at room temperature. By applying an external DC-voltage across a (011)-cut PMN-PT substrate, the ferroelectric polarization is re-oriented, which results in an anisotropic in-plane strain that transfers to the CoFeB thin film and changes its magnetic anisotropy $H_k$. The change in $H_k$ in-turn results in a 90° rotation of the magnetic easy axis for sufficiently high voltages. Simultaneously, the inverse effect is employed to read changes of the magnetic properties. Because the Piezoelectric (PE)/FerroMagnetic (FM) system is fully coupled, the change of magnetization in FM induces an elastic stress in the PE layer, which generates a piezoelectric potential in the system that can be used to readout the magnetic state of the FM layer. Our experimental results are in excellent qualitative agreement with a recently proposed, experimentally benchmarked equivalent circuit model that considers how magnetic properties are electrically controlled in such ME/PE heterostructure and how a back-voltage is generated due to changing magnetic properties in a self-consistent model.

Recently, so-called Magnetoelectric (ME) effects [1-4] in Piezoelectric (PE)/FerroMagnetic (FM) structures have attracted substantial research interest, since these may open a path towards controlling magnetic properties by means of an electric field – rather than currents – to achieve low energy dissipation in the writing process of magnetic memory cells [5-19]. As to the reading process, the Giant Magneto-Resistance (GMR) [18] or Tunneling Magneto-Resistance (TMR) [1] is usually utilized to readout the stored magnetic information. However, the readout energy based on magneto-resistive methods is not negligible compared with the extremely low writing energies. In addition, the integration of GMR or Magnetic Tunnel Junction (MTJ) stacks with piezoelectric substrates remains a challenge. This raises the question whether voltages that are used in case of the write operation could also be used in a reverse configuration to read the magnetic information in a more power efficient way. To demonstrate this idea, in this article, we have measured this reciprocal effect simultaneously and achieved a strain-mediated magnetoelectric write and read unit in a $Co_{60}Fe_{20}B_{20}$/ $Pb(Mg_{1/3}Nb_{2/3})_{0.7}Ti_{0.3}O_3$ heterostructure.

The write operation in the magnetic subsystem is accomplished by applying an external DC-voltage across a (011)-cut PMN-PT substrate. The re-orientation of the ferroelectric polarization results in an anisotropic in-plane strain [16], which transfers to the CoFeB thin film and changes its $H_k$ [1,16]. The change in $H_k$ in-turn results in a 90º rotation of the magnetic easy axis for high enough voltages. In the actual experiment, while applying a DC voltage to the PMN-PT substrate, an AC magnetic field is simultaneously applied in the hard axis direction of the CoFeB film to induce an AC magnetization change. As a feedback, this AC magnetization change in CoFeB introduces strain in the substrate and generates a readout voltage across the PMN-PT [3-4]. Since the AC magnetization change is dependent on the $H_k$ of the CoFeB which is being controlled by the DC voltage, the AC voltage generated (READ) across the PMN-PT is controlled by the external DC voltage applied (WRITE). Taking advantage of the coupling between the magnetic and electric effects, we have achieved both, electrical write and read operation in CoFeB/PMN-PT heterostructures by modulating the magnetization state of the CoFeB film with a DC voltage and reading out this state through the detected AC voltage across the PMN-PT substrate. Note that in contrast to a recent report [3] that also explores PE/FM systems, our device does not require an external magnetic field during the writing process to obtain two stable equilibrium states corresponding to the bit states "0" and "1", since the amorphous CoFeB film in our structure is magneto-crystalline isotropic [1,16].

Our results are in excellent qualitative agreement with a recently proposed equivalent circuit [4] that considers how the easy axis of the FM is changed due to an applied voltage and how a back-voltage ($v_m$) is generated in the PE due to this change in a self-consistent model. The theoretical and experimental results reported here suggest a different mode of operation from normal magnetic random-access memory (MRAM) technology where the bit states "0" and "1" are represented by a pseudo-magnetization $\mu \equiv m_x^2 - m_y^2$ rather than a net magnetization $m_x$, $m_y$ or $m_z$. The switching between "0" and "1" states by the write voltage is directly detected as a read signal. This means that the readout process can be performed without the need for an external magnetic field or tunnel junction. One concrete device application that one can imagine would utilize a PE/FM stack in series with a capacitor and would use a low-barrier nano-magnet to obtain an ultra-low power tunable random number generator [4]. This device could be a potential voltage controllable "p-bit" which can be used in a new type of probabilistic logic circuit [20-25]. Thus, magnetic field free magneto-electric read and write operations become feasible in a magnetic system, which makes this work highly relevant for voltage-controlled spintronics applications and the development of low-power and high-speed data storage technology in general [5-19].

## Results and Discussion

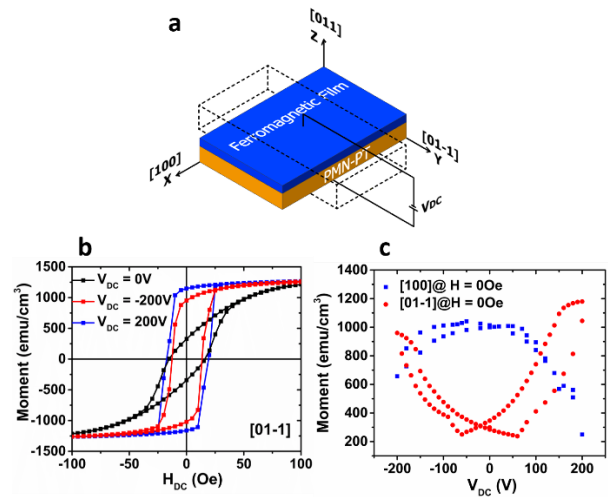

FIG. 1. (a) Schematic configuration of the

ferromagnetic/piezoelectric heterostructure and the piezo-response of PMN-PT under a sufficiently positive or negative DC voltage. (b) Magnetic hysteresis loops of an Au(100nm)/Ti(10nm)/(011)-cut PMN-PT(300µm)/ CoFeB(20nm)/Ta(5nm) heterostructure measured along the [01-1] direction under different voltages. (c) Voltage-control of magnetization measured along both [100] and [01-1] directions with H = 0Oe.

The magnetoelectric (ME) write operation has been achieved in a $Co_{60}Fe_{20}B_{20}$/ $Pb(Mg_{1/3}Nb_{2/3})_{0.7}Ti_{0.3}O_3$ heterostructure. Figure 1(a) shows the schematic configuration and the piezo-response of PMN-PT under a sufficiently positive or negative DC voltage. In all measurements, "positive DC voltage" means that the bottom electrode has a higher (more positive) potential than the top electrode. With DC voltages applied in the [011] crystalline direction, a strong in-plane anisotropic piezo-strain is induced due to the re-orientation of the ferroelectric polarization (P) [26-29] which transfers to the CoFeB thin film and provides an in-plane magnetic anisotropic field that changes the magnet's $H_k$ as discussed in Fig. A1. When the strain is sufficiently large, the result is a 90º rotation of the magnetic easy axis [1,16].

Voltage-controlled magnetization measurements were carried out in a Magnetic Property Measurement System (MPMS) at room temperature. Magnetic hysteresis loops of a PE/FM heterostructure consisting of a PMN-PT crystal and a 20nm CoFeB layer with in-plane magnetic anisotropy (IMA) measured along the [01-1] direction under different voltages are shown in Fig. 1(b). As discussed in Fig. A1, under large positive and negative voltages [case 2 and 5 in Fig. A1], there is a tensile strain in the [01-1] direction and a compressive strain in the [100] direction, which favors a magnetization along the [01-1] direction and results in a high remnant magnetization as evident from Fig. 1(b). When the voltage is removed, the compressive strain in the [01-1] direction and tensile strain in the [100] direction makes magnetization along the [01-1] direction harder and the remnant magnetization becomes low. Simultaneously, for the [100] direction, magnetization is easier for zero DC voltage and harder for large positive and negative voltages. The M-H loops of the [100] direction under different DC voltages are shown in Fig. A2. Figure 1(c) shows the change of remnant magnetization according to the applied DC voltages along both [100] and [01-1] directions, clearly indicating the complementary trend, i.e. decreasing moment towards large electric fields in the blue compared to increasing moment towards large electric fields in the red measurement case. Both exhibit a butterfly-like behavior, which is in consistent with the butterfly-like trend of induced in-plane strain in the PMN-PT [30].

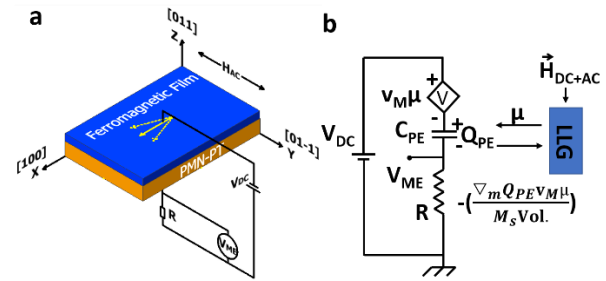

FIG. 2. (a) Experimental set-up and (b) an equivalent circuit model for the PE/FM heterostructure that captures the simultaneous read and write operation proposed in Ref [4].

Inversely, the ME read operation can also be achieved due to the coupling between the PE and FM layers. The change of magnetization induces deformation in the FM film due to the magnetostriction effect [31] which transmits to the PE substrate and results in a change of P due to the piezoelectric effect [29]. The variation of P during the reorientation of the magnetization gives rise to a magnetoelectric voltage defined as $V_{ME}$ across the heterostructure. The corresponding results on the ME read without applying an external voltage to the PE are discussed in Fig. A3.

In the following section, we will discuss the simultaneous experimental demonstration of ME write and read operation in the same PE/FM

sample. The heterostructure used in this measurement is CoFeB(40nm)/(011)-cut PMN-PT(300μm) with Ta and Ti/Au as the top and bottom electrodes respectively. The experimental set-up is shown in Fig. 2(a). Before moving on to the experimental results, we will discuss the measurement principle based on an equivalent circuit model illustrated in Fig. 2(b) [4]. The circuit has been experimentally benchmarked against the results reported in Ref [3].

The magnetization dynamics of the FM layer are calculated from the Landau-Lifshitz-Gilbert equation self-consistently which takes the stress-induced anisotropy field through the charge on the PE capacitor as an input and produces the pseudo-magnetization $\mu \equiv m_x^2 - m_y^2$ as an instantaneous output. The circuit equation is given as:

$$V_{DC} = \frac{\partial E_M}{\partial Q_{PE}} + \frac{Q_{PE}}{C_{PE}} + R\frac{dQ_{PE}}{dt}$$

where $E_M$ is the magnetic energy, $C_{PE}$ is the capacitance of the PE material, $Q_{PE}$ is the charge on the PE capacitor and R is a resistor connected in series with the PE/FM heterostructure. The write operation is accomplished by an effective field:

$$\vec{H}_{ME} = -\frac{\nabla_m E_M}{M_S Vol.}$$

where $\nabla_m$ is the gradient operator with respect to the magnetization directions $m_i$. $M_S$ and Vol. are the saturation magnetization and the volume of the magnetic film respectively. The read operation is accomplished due to a changing "back-voltage" through the pseudo-magnetization μ, $V_{ME\_tot} = \frac{\partial E_M}{\partial Q_{PE}} = v_m \mu$, where $\mu = m_x^2 - m_y^2$ is the pseudo-magnetization, $v_m$ is the back-voltage constant and $E_M = Q_{PE} v_m \mu$. To measure the magnetoelectric voltage using a lock-in amplifier, a resistor R with a resistance value comparable to the impedance of the piezoelectric capacitor is connected in series with the heterostructure. $V_{ME}$ measured across R is used to read the magnetic information of the FM layer. MATLAB analytical simulations on simultaneous ME write and read operations are discussed in Fig. A4.

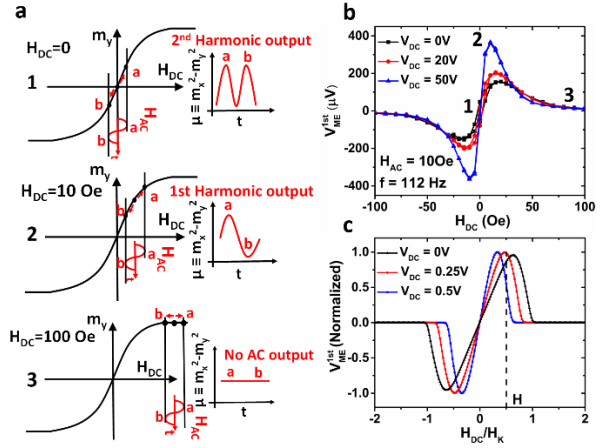

FIG. 3. (a) Schematic diagrams of the magnetization oscillation in the presence of an AC magnetic field applied on top of different DC magnetic fields in the [01-1] direction and the corresponding AC outputs. Case 1-3 illustrate the situations when $H_{DC}$ = 0Oe, 10Oe and 100Oe respectively. (b) Experimental results on $V_{ME}^{1st}$ measured across R with an AC magnetic field superimposed on a DC magnetic field along the [01-1] direction when $V_{DC}$ = 0V, 20V and 50V. States 1-3 correspond to the cases 1-3 in (a). (c) Simulation results of the normalized in-phase component of the AC voltage $V_{ME}^{1st}$ as a function of DC magnetic fields that is read out from the circuit in Fig. 2(b). Without loss of generality, the following parameters are used in the numerical circuit: $H_K$=25 Oe, $M_s$=300 emu/cc, Vol. = 5e-13 cm³, damping coefficient α=0.1, demagnetization field $H_D$ is assumed to be 100 $H_K$. The piezoelectric capacitance $C_{PE}$ is chosen to be 7.5 pF, and the load resistance is 0.1 Ω. Note that for the normalized $V_{ME}^{1st}$ the actual value of the back-voltage $v_m$ does not change the simulation results.

Experimentally, $V_{ME}$ is measured in the presence of an AC magnetic field applied on top of a DC magnetic field in the [01-1] direction. The amplitude of the AC magnetic field is 10Oe and the frequency is 112Hz. Before discussing the experimental results, Fig. 3(a) shows how the magnetization ($m_y$) oscillates in the presence of an AC magnetic field applied on top of different DC magnetic fields in the [01-1] direction and the corresponding AC outputs. Based on the analysis in Fig. 1, the hard axis lies in the [01-1] direction when $V_{DC}$ = 0V. Thus, the magnetization in the y-direction ($m_y$) is approximately a sigmoidal function (tan-hyperbolic) of the magnetic field applied in the y-direction. In case 1 of Fig. 3(a), when $H_{DC}$=0Oe, applying $H_{AC}$, results in $m_y$ and thus the pseudo-magnetization $\mu = m_x^2 - m_y^2$ to oscillate between points a and b. The oscillation frequency of $\mu = m_x^2 - m_y^2$ is twice that of $H_{AC}$ with $\mu(a) = \mu(b)$. According to Eq. (A4) and (A5) in the appendix section, the 1st harmonic of $V_{ME}$ is thus zero and the AC output in case 1 only has contributions to the 2nd harmonic of $V_{ME}$ (see also Fig. A3(e) and (h)). However, for a finite $H_{DC}$ (e.g. 10 Oe), the pseudo-magnetization oscillates with the same frequency as $H_{AC}$. Thus, $\mu = m_x^2 - m_y^2$ is different for points a and b and a finite 1st harmonic output signal is generated as shown in case 2 of Fig. 3(a). Last, for $H_{DC}$=100Oe, $m_y$ is in the saturation region and hence neither the magnetization nor the pseudo-magnetization oscillate with $H_{AC}$, which results in a zero AC output signal. The experimental results in Fig. 3(b) follow this trend from case 1 to 2 and 3. Note that for a given $H_{AC}$, the oscillation amplitude of $m_y$ and $\mu$ is dependent of the slope of the $m_y$-$H_{DC}$, i.e. $H_k$. Since $V_{DC}$ applied across the PE-ME structure modifies $H_k$ of the magnetic film, the AC output signal is expected to be different for different $V_{DC}$-values. Positive $V_{DC}$-values result in steeper $m_y$-$H_{DC}$ curves as shown in Fig. A4(b) and (c) and hence increase the oscillation amplitude of $\mu = m_x^2 - m_y^2$ for a given $H_{AC}$, which results in a higher peak amplitude of $V_{ME}$.

Figure 3(b) shows experimental results on the relationship between the first harmonic $V_{ME}$ i.e. $V_{ME}^{1st}$ and $H_{DC}$ when $V_{DC}$= 0V, 20V and 50V. The experimental results clearly show that both, the peak amplitude ($V_P$) and the position of the peak ($H_P$) are different for different $V_{DC}$ following a trend: $V_{P\_50V} > V_{P\_20V} > V_{P\_0V}$ and $H_{P\_50V} < H_{P\_20V} < H_{P\_0V}$, which corresponds to the different $H_k$ of the CoFeB film caused by $V_{DC}$. The actual M-H loops of the CoFeB film that correspond to Fig. 3(b) are shown in Fig. A5(a).

Furthermore, the experimental results are described qualitatively by theoretical simulations using the equivalent circuit shown in Fig. 2. Our equivalent circuit SPICE model calculates a time dependent magnetization and voltage, allowing us to replicate even dynamics of the experimental measurement self-consistently. According to the M-H loops shown in Fig. A6, when $V_{DC}$ increases from 0V to 0.5V, the magnetization in the [01-1] directions becomes easier which is consistent with the analysis in Fig. 1. From Eq. (A4) in the appendix section, the first harmonic in-phase load voltage $V_{ME}^{1st}$ is calculated using our SPICE-model as a function of $H_{DC}$ at different $V_{DC}$ inputs. The corresponding results are shown in Fig. 3(c). The trends for a change in peak amplitude and peak position are qualitatively the same as in the experiment, i.e. higher $V_P$ and smaller $H_P$ for smaller $H_k$.

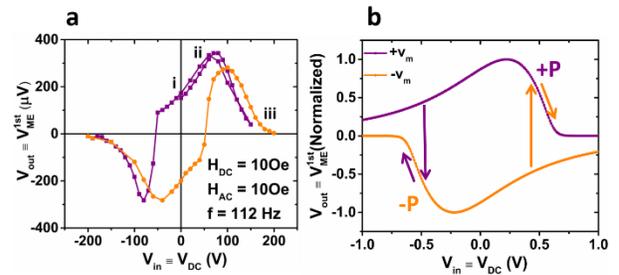

FIG. 4. (a) Experimental results on the relation between $V_{in} \equiv V_{DC}$ and $V_{out} \equiv V_{ME}^{1st}$ when $H_{DC}$ = 10Oe, $H_{AC}$ = 10Oe and the frequency of the AC field is 112Hz. The purple and orange lines represent scanning $V_{in} \equiv V_{DC}$ from +200V to -200V and from -200V to +200V respectively. (b) The purple and orange lines are simulation results on $V_{ME}^{1st}$ as a function of $V_{DC}$ at a fixed DC magnetic field corresponding to 0.5 $H_{DC}/H_K$ that is read out from the circuit in Fig. 2(b) when the back-voltage constant is +$v_m$ and -$v_m$ respectively. Since our simulation does not capture a voltage controlled ferroelectric polarization switching, positive and negative $v_m$ values had to be used in the simulation. The arrows indicate how $V_{out}$ versus $V_{in}$ would behave once polarization switching is considered.

When envisioning a device application, the idea is to use voltage-to-spin conversion for the write operation and spin-to-voltage conversion for the read. Thus, the types of device characteristics that are desirable would have a voltage input $V_{in}$ and voltage output $V_{out}$, with the magnetic information mediating between the two. To show how an input DC voltage, can control indeed in our devices a $V_{ME}$ output voltage, we have fixed $H_{DC}$ at 10Oe and have swept $V_{in}$ between -200V and 200 V. The experimental and spice-simulation results that show the relation between $V_{in} \equiv V_{DC}$ and $V_{out} \equiv V_{ME}^{1st}$ are displayed in Fig. 4(a) and (b). As explained in the previous paragraph, both the peak amplitude and position of $V_{ME}^{1st}$ ($H_{DC}$) are different for different $V_{DC}$ due to the change of $H_k$ in the CoFeB film. As shown in Fig. 3(c), when $H_{DC}$ is fixed at "H" which corresponds to 0.5 $H_{DC}/H_K$, $V_{ME}^{1st}$ ($V_{DC}$ = 0.5V) < $V_{ME}^{1st}$ ($V_{DC}$ = 0V)< $V_{ME}^{1st}$ ($V_{DC}$ = 0.25V). Since the states (i) through (iii) in Fig. 4(a) correspond to the states of $V_{DC}$ = 0V, 0.25V and 0.5V respectively in Fig. 3(c), one obtains $V_{ME}^{1st}$ (iii) < $V_{ME}^{1st}$ (i)< $V_{ME}^{1st}$ (ii). The abrupt switch in both the purple and orange lines around +/- 50V corresponds to the piezoelectric polarization (P) switching when scanning $V_{DC}$. The relation between the piezoelectric polarization P and $V_{ME}^{1st}$ is shown in Fig. A5(b) indicating $V_{ME}^{1st}$ (P) = -$V_{ME}^{1st}$ (-P).

Figure 4(b) illustrates simulation results on the normalized $V_{ME}^{1st}$ as a function of $V_{DC}$ at a fixed DC magnetic field corresponding to 0.5 $H_{DC}/H_K$ that is read out from the circuit in Fig. 2(b). Note that to capture the hysteretic nature of $V_{ME}^{1st}$ versus $V_{DC}$ in the simulation, the coefficient $v_m$ is made negative when the polarization switched. Compared with Fig. 4(a), all the experimental results are in strong qualitative agreement with those predicted by the circuit model.

Next, to extract the back-voltage constant in our experimental heterostructure, we performed a quantitative benchmarking of our results. Experimentally, when $V_{in}$ = 50V, $V_{ME}^{1st}$ = 290μV with $H_{AC}$ = 10Oe and $H_k$ = 70Oe read from the M-H loop shown in Fig. A5(a). SPICE simulations were performed after considering the circuit loss factor $\frac{|jwRC|}{|jwRC+1|} = 0.37$ in the experimental set-up and all other experimental parameters. The coefficient $v_m$ was calculated to be 5.5mV according to Eq. (A4) in the appendix section to match $V_{ME}^{1st}$ = 290μV when $\frac{H_{AC}}{H_k} = \frac{10Oe}{70Oe}$. $v_m$ is a measure of the particular coupling strength between the chosen PE and FM materials. A theoretical value can be calculated from $v_m = \frac{Bdt_{FM}}{2\varepsilon}$ using the thickness of the FM layer $t_{FM}$ = 40nm, the magneto-elastic constant B = -4MPa [32], the net PE constant d = $d_{31} - d_{32}$ = 4500 pm/V, and the permittivity of the PE layer ε = 600$\varepsilon_0$ [4]. The estimated $v_m$ = 70mV is about one order of magnitude higher than the one extracted from the experimental results. As a point of reference, from the experimental results in Ref [3] on [N*(TbCo2/FeCo)]/PMN-PT a $v_m$ of 37mV can be extracted according to the analysis in Ref [4]. Using $v_m = \frac{Bdt_{FM}}{2\varepsilon}$ on the other hand, one finds $v_m$= 49mV. While in both cases the experimentally extracted $v_m$ is lower than the calculated one, we believe that the larger discrepancy in our case is likely a result of the non-ideal strain transfer between CoFeB and PMN-PT, which will require further studies in the future.

In conclusion, taking advantage of the coupling between magnetic and electric effects, we have achieved both, electrical write and read operation in CoFeB/PMN-PT heterostructures simultaneously. Since the strain induced in the PMN-PT is proportional to the electrical field applied across the structure, a 30nm film would allow decreasing the write voltage $V_{in}$ to ±20mV, making our approach attractive for various magnetic device applications. Moreover, our experimental results are qualitatively consistent with the theoretical simulations of an equivalent circuit reported in Ref [4]. In this work, a novel magnetic operation mode is proposed where a pseudo-magnetization $\mu = m_x^2 - m_y^2$ rather than a net magnetization $m_x$, $m_y$ or $m_z$ is used as the bit states in MRAM technology. The experimental and theoretical work reported here promises the feasibility of magnetic field free magneto-electric read and write operations in a magnetic system by utilizing a PE/ME stack in series with a capacitor and using a low-barrier nano-magnet to obtain an ultra-low power tunable random number generator [4] which is significant for the development of voltage-controlled spintronics and low power, high-speed data storage technology.

## Methods

SPICE simulation:

As in the experiment, first a constant DC field is applied to the FM in the in-plane hard axis [01-1] direction during the entirety of the transient simulation (10ns for the present case). After the first 5ns to allow for the magnet dynamics to reach a steady state, an AC magnetic field with a 500MHz frequency and amplitude of $0.1H_K$ is applied along the same direction.

AUTHOR INFORMATION

Corresponding Author

*E-mail: shen168@purdue.edu; vostwal@purdue.edu

Author contributions

T.S. worked on the device fabrication, characterization and data analysis; V.O. worked on characterization, data analysis and MATLAB analytical simulation. K.Y.C. worked on data analysis and HSPICE simulation of the circuit model. J.A. analyzed the data and oversaw the planning and execution of the project; T.S. wrote the manuscript.

Notes

The authors declare no competing financial interest.

## Acknowledgements

This work was supported in part by the Center for Probabilistic Spin Logic for Low-Energy Boolean and Non-Boolean Computing (CAPSL), one of the Nanoelectronic Computing Research (nCORE) Centers as task 2759.003 and 2759.004, a Semiconductor Research Corporation (SRC) program sponsored by the NSF through CCF 1739635. The authors thank the staff at the Birck Nanotechnology center for their technical support. T. Shen is thankful for the help from Neil Dilley for discussions about experimental measurements.

## APPENDIX A: WORKING PRINCIPLE OF (011)-CUT PMN-PT

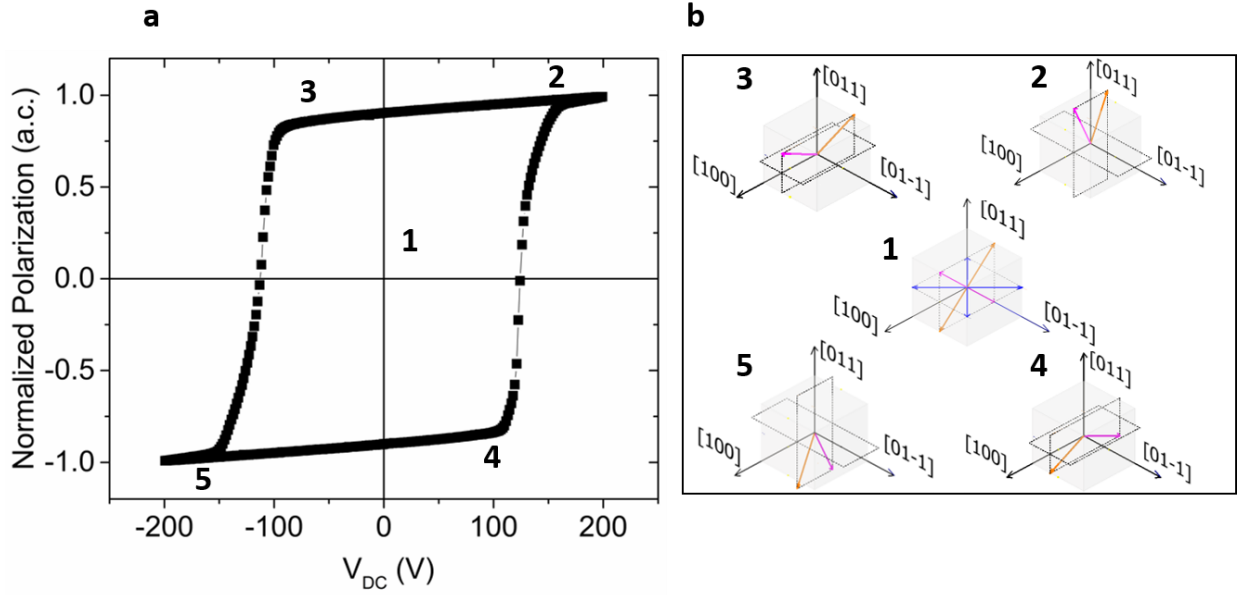

FIG. A1. (a) Experimentally measured ferroelectric polarization versus DC voltage properties of the PMN-PT. (b) The unit cell and spontaneous ferroelectric polarizations of (011)-cut PMN-PT correspond to different cases labeled by 1-5 in (a).

Single crystal PMN-PT with (011) orientation was employed in this work as the piezoelectric substrate for achieving large ME effects due to its large piezoelectric coefficients with $d_{31} \sim -3100$ pC/N along the [100] direction and $d_{32} \sim 1400$ pC/N along the [01-1] direction [26-29]. With DC voltages applied in the [011] crystalline direction, a strong in-plane anisotropic piezo-strain is induced due to the re-orientation of the ferroelectric polarization (P) [26-29]. As shown in Fig. A1(a), the PMN-PT exhibits a typical square-shaped ferroelectric hysteresis behavior at room temperature. The inserts one through five in Fig. A1(b) represent the unit cell and spontaneous ferroelectric polarizations of [011] cut PMN-PT under different DC voltages in accordance with the 1-5 states labeled in Fig. A1(a) respectively.

For PMN-PT with rhombohedral (R) phase, the spontaneous polarizations are along the [111] directions, which means the diagonals of the (011) and (01-1) plane in the (011)-cut crystal. Before poling, the polarization orientations are randomly distributed in the eight possible directions and the total polarization P = 0, as shown in case 1. In this situation, there is no net strain in the PMN-PT. When the PMN-PT is poled along the [011] direction by a large positive voltage, a net polarization in the [011] direction occurs as a result of the only two remaining polarization vectors – yellow and pink in Fig. A1(b). If compared to the two initial polarization vectors with a positive component in the [011] direction as displayed in case 1, case 2 exhibits a larger net polarization of these two vectors, leading to a compressive strain in the [100] direction and a tensile strain in the [01-1] direction. After removing the electric field and applying a small negative voltage that is not large enough to switch the polarization to the other direction, the net polarization is still in the [011] direction but its magnitude is smaller than the sum of the two initial polarization vectors (pink and yellow) with positive component in the [011] direction. Thus, if compared with the un-poled state, the crystal is elongated in the [100] direction and compressed in the [01-1] direction as shown in insert 3. Similarly, when the PMN-PT is poled along the negative [011] direction by a sufficiently large negative voltage, a net polarization that points in the negative [011] direction is obtained as shown in insert 5. This state is characterized, like case 2, by a compressive strain in the [100] direction and a tensile strain in the [01-1] direction. Finally, by removing the negative voltage and applying a small positive voltage, state 4 is reached that is characterized by the same situation as case 3. The anisotropic in-plane strain that can in this way be produced by applying a voltage to the PMN-PT crystal transfers to the CoFeB thin film and provides an in-plane magnetic anisotropic field that changes the magnet's $H_k$.

## APPENDIX B: ME WRITE OPERATION

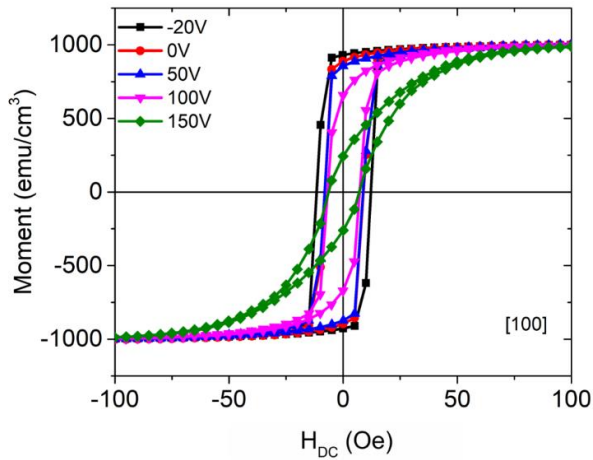

FIG. A2. M-H loops of an Au(100nm)/Ti(10nm)/(011)-cut PMN-PT(300μm)/CoFeB(10nm)/Ta(5nm) heterostructure measured along the [100] direction under different voltages.

The M-H loops of the CoFeB/PMN-PT stack along the [100] direction of an Au(100nm)/Ti(10nm)/(011)-cut PMN-PT(300μm)/CoFeB(10nm)/Ta(5nm) heterostructure under different DC voltages are shown in Fig. A2. The measurements were carried out in a Magnetic Property Measurement System (MPMS) with in situ DC voltages applied as in the configuration shown in Fig. 1(a). Positive DC voltages increase the magnetic anisotropy along the [100] direction while small negative DC voltages decrease it. The results are consistent with the description of strain formation as presented in the main text and complementary to the results shown in Fig. 1(b). Positive voltages produce a compressive strain in the [100] direction and a tensile strain in the [01-1] direction, while small negative voltages lead to a tensile strain in the [100] direction and a compressive strain in the [01-1] direction. When the anisotropic piezo-strain is large enough, the easy axis of the CoFeB film rotates by 90°.

## APPENDIX C: ME READ OPERATION

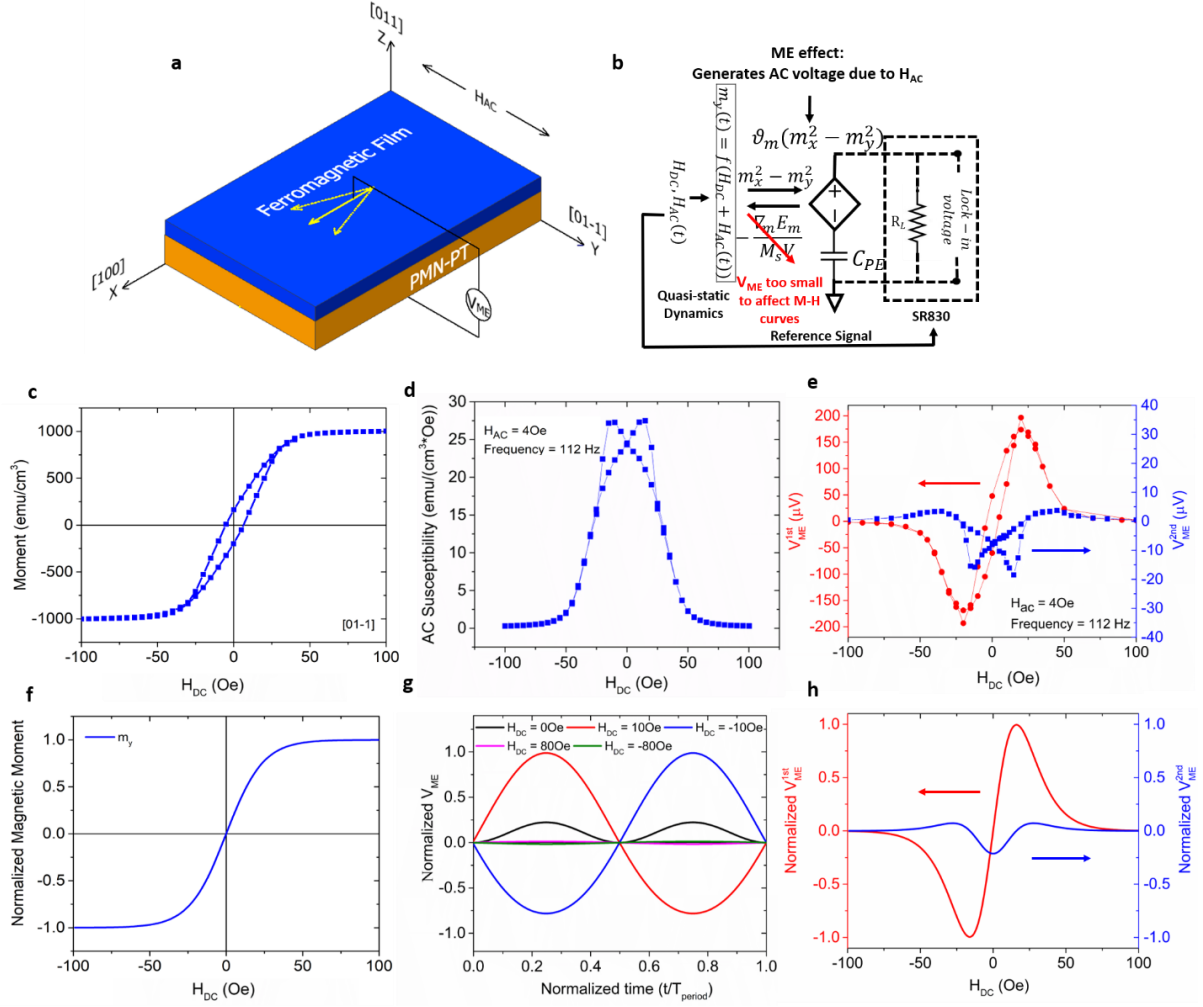

FIG. A3. (a) Experimental set-up and (b) circuit model of the ME read operation. Experimental results on (c) the magnetic hysteresis in the [01-1] direction without external DC voltage; (d) the AC magnetic susceptibility with $H_{AC}$ = 4Oe and f = 112 Hz applied in the [01-1] direction; and (e) the resulting $V_{ME}^{1st}$ and $V_{ME}^{2nd}$. Analytical results on (f) the M-H loop in the [01-1] direction without external DC voltages; (g) Normalized total $V_{ME}$ within one period of an AC magnetic field with a magnitude of $H_{AC}$ = 4Oe and frequency of 112Hz in the Y direction at different DC magnetic fields; and (h) the corresponding $V_{ME}^{1st}$ and $V_{ME}^{2nd}$.

As discussed in the main text, the ME effect can be used to read the magnetic state in the CoFeB film due to the magnetostriction effect [31] and the piezoelectric effect [29]. Experimentally, we have carried out the ME read operation in a PE/FM heterostructure with a 40nm CoFeB film deposited on a poled 300µm thick (011)-cut PMN-PT with Ta and Ti/Au as the top and bottom electrodes respectively. Figure A3(a) and (b) show the experimental set-up and circuit model respectively. According to the analysis in Fig. 1, the hard axis of the CoFeB film is along the [01-1] direction when $V_{DC}$ = 0V, which is confirmed further by the "sloppy" magnetic hysteresis from another device shown in Fig. A3(c). By applying an AC magnetic field $H_{AC}$ with amplitude of 4Oe and frequency of 112Hz on top of a DC magnetic field in the [01-1] direction, an AC magnetization change is induced in the way shown by the yellow arrows in Fig. A3(a). Figure A3(d) shows the resulting AC magnetic susceptibility, i.e. the slope of the M-H curve shown in Fig. A3(c). The magnetoelectric voltage $V_{ME}$ induced by the magnetization variation is measured across the PMN-PT by a lock-in amplifier SR830. The red and blue curves in Fig. A3(e) are the first and second harmonic, i.e. $V_{ME}^{1st}$ and $V_{ME}^{2nd}$ respectively.

To explain the complicated experimental results, we analytically replicated the measurements based on the circuit model proposed in Ref [4] using MATLAB. The magnetization in the y-direction ($m_y$) is assumed to be a sigmoidal function (tan-hyperbolic) of the magnetic field ($H_y$) applied in the y-direction, closely matching the experimentally obtained M-H curve shown in Fig. A3(c). As in the experiments, the total magnetic field ($H_y$) at time (t) is $H_{DC} + H_{AC} * \sin(wt)$ where ω is the modulation frequency of the AC magnetic field with an amplitude $H_{AC}$, and $H_{DC}$ is the DC field. At a given time t, the magnetization is calculated as:

$$m_y(t) = f(H_{DC} + H_{AC} * \sin(wt)) \quad (A1)$$

where f is the tan-hyperbolic function defined above. Since the total magnetization $m_x^2(t) + m_y^2(t)$ remains 1 for all applied magnetic fields in the X-Y plane, we define the pseudo-magnetization as a function of $t$ for different $H_{dc}$ ranging from -100 to +100 Oe as:

$$\mu \equiv m_x^2(t) - m_y^2(t) = 1 - 2 * m_y^2(t) \\ = 1 - 2 * f(H_{DC} + H_{AC} * \sin(wt)) \quad (A2)$$

According to the analysis in Ref [4], the voltage generated due to the magnetoelectric effect is calculated by

$$V_{ME}(t) = v_m * (m_x^2(t) - m_y^2(t)) \quad (A3)$$

where $v_m$ is the back-voltage constant determined by the coupling strength between the PE and FM layers. Figure A3(g) shows the results of $V_{ME}$ as a function of time for HAC = 4Oe with different $H_{DC}$. For an AC magnetic field of frequency 2π/w, the 1st and 2nd harmonic of $V_{ME}$ is extracted using

$$V_{ME}^{1st} = \frac{1}{T} \int_0^T V_{ME}(t) * \sin(wt) dt \quad (A4)$$

$$V_{ME}^{2nd} = \frac{1}{T} \int_0^T V_{ME}(t) * \sin(2wt) dt \quad (A5)$$

for different $H_{DC}$ and the corresponding results are plotted in Fig. A3(h).

As apparent, both the 1st and 2nd harmonics of the experimental and analytical results show a good qualitative match, which indicates that the simple analytical expressions from above can be used to explain the experimentally observed ME effect. Note that due to the hysteresis in the experimental M-H curve in Fig. A3(c) that might be a result of the hard axis not being perfectly aligned with the [01-1] crystalline direction, both harmonics curves are shifted along the x-axis depending on the direction of the $H_{DC}$ sweep. It is worth to mention that in the analysis from above, $V_{ME}(t)$ is assumed to be too small to affect the magnetic properties of the CoFeB film. Without this assumption, one has to self-consistently solve problem, where the change in the magnetic properties is generating $V_{ME}$ and the generated $V_{ME}$ is affecting the magnetic properties. This approach has been used in the main text by utilizing the circuit shown in Fig. 2(b). The above assumption of $V_{ME}(t)$ being small is valid for the measurements in Fig. A3 since to change the M-H curves, at least tens of volts are required as shown in Fig. 1, while the voltage -$V_{ME}$ - generated by the AC magnetic field is only in the range of a few hundred microvolts.

# APPENDIX D: SIMULTANEOUS ME WRITE AND READ OPERATIONS

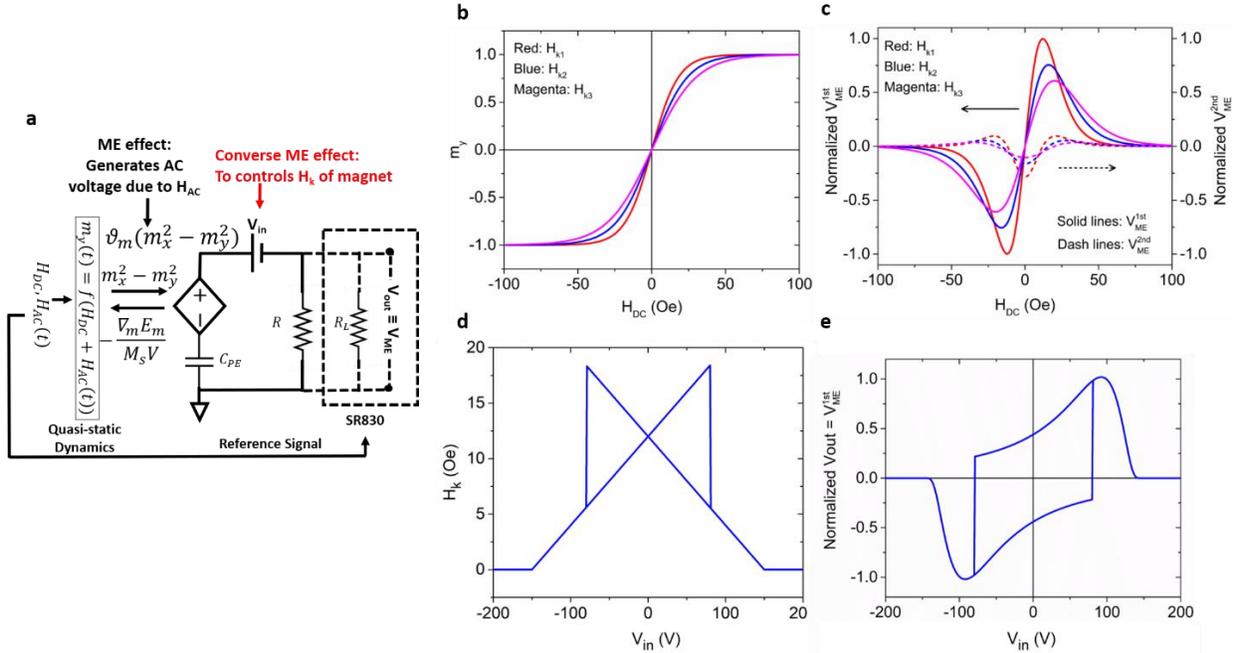

FIG. A4. (a) Circuit model of the simultaneous ME write and read operations. (b) Theoretical M-H loops of the [01-1] direction for different $H_k$ with $H_{k1} < H_{k2} < H_{k3}$; (c) The first and second harmonic of $V_{ME}$, i.e. $V_{ME}^{1st}$ and $V_{ME}^{2nd}$ corresponding to different $H_k$ values (d) The relation between $H_k$ of the magnetic film and the external DC voltage used in the simulation. (e) The relation between $V_{in}$ and $V_{out} \equiv V_{ME}^{1st}$ when $H_{DC}$ = 10Oe, $H_{AC}$ = 10Oe and the frequency of the AC field is 112Hz.

Figure A4 shows simulation results on the simultaneous ME write and read operations using the same method as discussed above in Appendix C. The circuit model is shown in Fig. A4(a) – the magnetic state of the FM layer is controlled by the input voltage $V_{in}$ and read out by the induced magnetoelectric voltage. To measure the magnetoelectric voltage using a lock-in amplifier SR830, a resistor R with a resistance value similar to the impedance of the piezoelectric capacitor is connected in series in the circuit. $V_{out} \equiv V_{ME}^{1st}$ measured across R is used to read the magnetic information in the FM layer. Figure A4(b) is the theoretical M-H loops in the Y direction for different $H_k$ ($H_{k1} < H_{k2} < H_{k3}$) in line with the experimental results on ME write operation discussed in Fig. 1, however without hysteresis. The corresponding first and second harmonic of $V_{ME}$ i.e. $V_{ME}^{1st}$ and $V_{ME}^{2nd}$ are shown in Fig. A4(c). Figure A4(d) shows the relation between $H_k$ of the magnetic film and the external DC voltages used in the simulation based on the analysis in Fig. 1, i.e. $H_k$ in the Y axis decreases under large positive and negative voltages because the tensile strain in the Y direction and compressive strain in the X direction favors a magnetization along the Y direction. The butterfly-like hysteresis is in consistent with the butterfly-like trend of induced in-plane strain in the PMN-PT [30]. Accordingly, the relation between $V_{in}$ and $V_{out} \equiv V_{ME}^{1st}$ is illustrated in Fig. A4(e) when $H_{DC}$ = 10Oe, $H_{AC}$ = 10Oe and the frequency of the AC field is 112Hz. This set of results qualitatively matches with both the experimental and SPICE simulation results shown in Fig. 3 and Fig. 4.

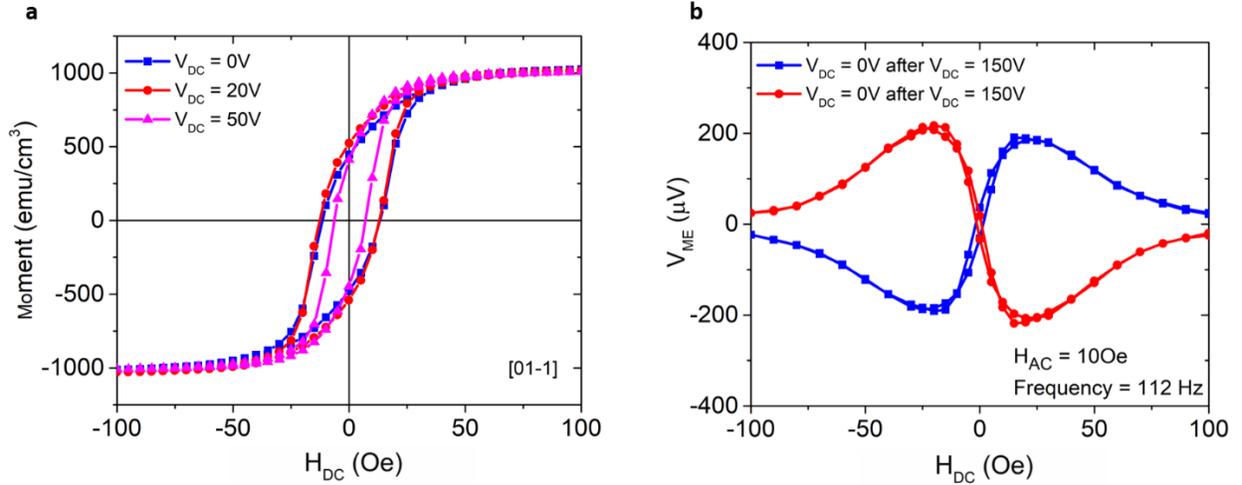

FIG. A5. (a) M-H loops of the Au(100nm)/Ti(10nm)/(011)-cut PMN-PT(300μm)/CoFeB(40nm)/Ta(5nm) heterostructure discussed in Fig. 3 along the [01-1] when $V_{DC}$ = 0V, 20V and 50V. (b) The relation between $V_{ME}^{1st}$ and $H_{DC}$ when $V_{DC}$ = 0V after the device had been conditioned by applying a $V_{DC}$ = 150V and -150V respectively.

Experimentally, as discussed in the main text and shown in Fig. 3 and Fig. 4, we have achieved simultaneous ME write and read in an Au(100nm)/Ti(10nm)/(011)-cut PMN-PT/CoFeB(40nm)/Ta(5nm) heterostructure. In Fig. A5, the blue, red and magenta curves are the M-H loop of the CoFeB film measured along the [01-1] direction when $V_{DC}$ = 0V, 20V and 50V respectively. From the slopes of these three curves, the $H_k$ of the CoFeB film decreases when increasing $V_{DC}$ confirming the results from Fig. 1(b).

To study the relationship between the piezoelectric polarization P and $V_{ME}$, the PMN-PT was poled in both positive and negative [011] direction with large $V_{DC}$. In Fig. A5(b), the blue and red lines are $V_{ME}^{1st}$ when $V_{DC}$ = 0V after applying +150V and -150V respectively. The $V_{ME}^{1st}$ in these two cases has the same magnitude but opposite signs. In other words, $V_{ME}$(P) = $-V_{ME}$(-P). This indicates that the hysteresis between the purple and orange lines in Fig. 4 is caused by the polarization hysteresis of the PMN-PT.

## APPENDIX E: SPICE SIMULATION ON ME WRITE AND READ OPERATIONS

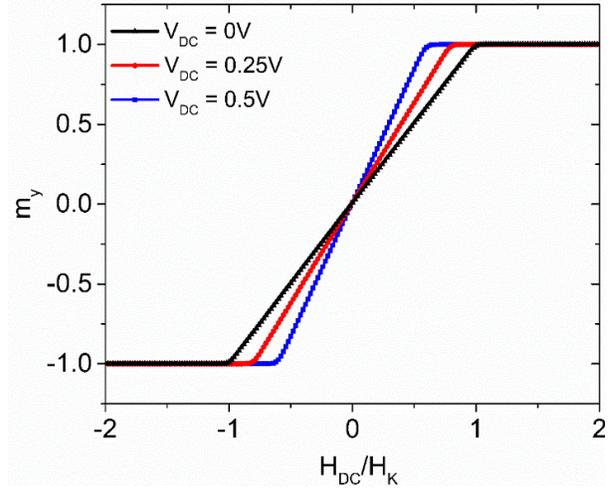

FIG. A6. Average in-plane hard axis magnetization $m_y$ as a function of DC magnetic field in the y-direction at different DC voltages ($V_{DC}$).

Taking advantage of a circuit simulator (HSPICE) the experimental results are repeated as a numerical experiment self-consistently as explained in the main text. The detailed working principles of the simulation are described in the main text. Figure A6 shows the average magnetization at different input voltages, verifying the easy-axis anisotropy modulation by different input voltages which qualitatively agrees well with the experimental results in Fig. A5(a). Thus, the pseudo-magnetization $\mu \equiv m_x^2 - m_y^2$ rather than a net magnetization $m_x$, $m_y$ or $m_z$ can be used as a novel magnetic operation mode in MRAM technology, which opens new possibilities for magnetic writing and reading.